\documentclass[11pt]{article}
\usepackage{moriond,epsfig}

\def\Journal#1#2#3#4{{#1} {\bf #2}, #3 (#4)}

\def\be{\begin{equation}}
\def\ee{\end{equation}}
\def\bea{\begin{eqnarray}}
\def\eea{\end{eqnarray}}

\begin{document}
\vspace*{4cm}
\title{Hadron spectroscopy with COMPASS -- First results}
\vspace{-6.3cm}
\author{Frank Nerling \\}
\address{Physikalisches Institut, Universit\"at Freiburg,\\ Hermann-Herder-Str. 3, D-79104 Freiburg, Germany \\
nerling@cern.ch}
\author{\vspace{-0.3cm}for the COMPASS collaboration}
\maketitle
\vspace{0.3cm}
\abstracts{
The COMPASS experiment at CERN is dedicated to light hadron spectroscopy with emphasis
on the detection of new states, in particular the search for spin exotic states and glueballs.
After a short pilot run in 2004 (190 GeV/c $\pi^{-}$ beam, Pb target) showing significant
production strength of an exotic $J^{PC}=1^{-+}$ state at 1.66\,GeV/${\rm c^2}$, we have collected data
with 190 GeV/c hadron beams on a liquid hydrogen target in 2008/09.
The spectrometer features good coverage by calorimetry and our data provide excellent
opportunity for simultaneous observation of new states in different decay modes. 
The diffractively produced $(3\pi)^{-}$ system for example can be studied in both modes 
$\pi^{-}p \rightarrow \pi^{-}\pi^{+}\pi^{-}p$ and $\pi^{-}~p \rightarrow \pi^{-}\pi^{0}\pi^{0}~p$.
Reconstruction of charged and neutral mode rely on completely different parts of the apparatus. Observation of new states in 
both modes provides important checks. The results on diffractive dissociation into 3$\pi$ final states 
from the 2004 data recently published are discussed as well as the first comparison of neutral vs. charged 
mode based on a first partial wave analysis of 2008 data. 
}
\section{Introduction}
\label{sec.intro}
\vspace{-0.3cm}
The COMPASS fixed-target experiment~\cite{compass:2007} at CERN SPS is dedicated to the study of structure and 
dynamics of hadrons, both aspects of non-perturbative QCD. In a first phase (2002-2007) COMPASS studied the nucleon 
spin structure by deep inelastic scattering of 160\,GeV/c polarised muons off polarised $^6$LiD and NH$_3$ targets. 
During a second phase in 2008/09, measurements with 190\,GeV/c hadron beams on a liquid hydrogen and nuclear targets 
were performed, mainly dedicated to hadron spectroscopy. 
Of particular interest is the search for non-$q\bar{q}$ mesons, like hybrids, glueballs or tetra-quarks, which do 
not fit into the Constituent Quark Model (CQM), but are allowed within QCD. Such objects, however, would mix with 
ordinary $q\bar{q}$ states of same $J^{PC}$ quantum numbers, making it difficult to disentangle the different contributions. 
The observation of spin-exotic states with quantum numbers forbidden in the CQM, like $J^{PC}= 0^{--},~0^{+-},~1^{-+},~...~$, 
would provide clear evidence for physics beyond the CQM and thus a fundamental confirmation of QCD.
The lowest mass hybrid candidate, means a meson consisting of a colour octet $q\bar{q}$ pair neutralised in colour by an excited gluon, 
is predicted \cite{Morningstar:2004} to have exotic $J^{PC}=1^{-+}$, and a mass between 1.3 and 2.2\,GeV/c$^2$.  
Two experimental candidates for a $1^{-+}$ hybrid in the light quark sector have been reported in the past, the $\pi_1(1400)$ 
and the $\pi_1(1600)$. 
The latter was observed in diffractive production by different experiments, decaying -- among others, like $\eta'\pi$, $f_1(1285)\pi$ and 
$\omega\pi\pi$ -- into $\rho\pi$ \cite{Adams:1998,Khokhlov:2000}. 
Especially the resonance nature of both candidates, is still heavily disputed \cite{Amelin:2005,Dzierba:2006}.  
To improve our understanding, new experiments have to be performed at high statistics, extending the 
spectrum to masses beyond 2.2\,GeV/c$^2$. COMPASS has just started to shed new light on the puzzle~\cite{Alekseev:2009a}, as discussed in this paper.
%
%
\vspace{-0.3cm}
\section{The COMPASS experiment}
\label{sec.spectro}
\begin{figure}[tp!]
\vspace{-0.6cm}
  \begin{minipage}[h]{.59\textwidth}
    \begin{center}
     \includegraphics[clip, trim= 45 60 55 80,width=1.0\linewidth]{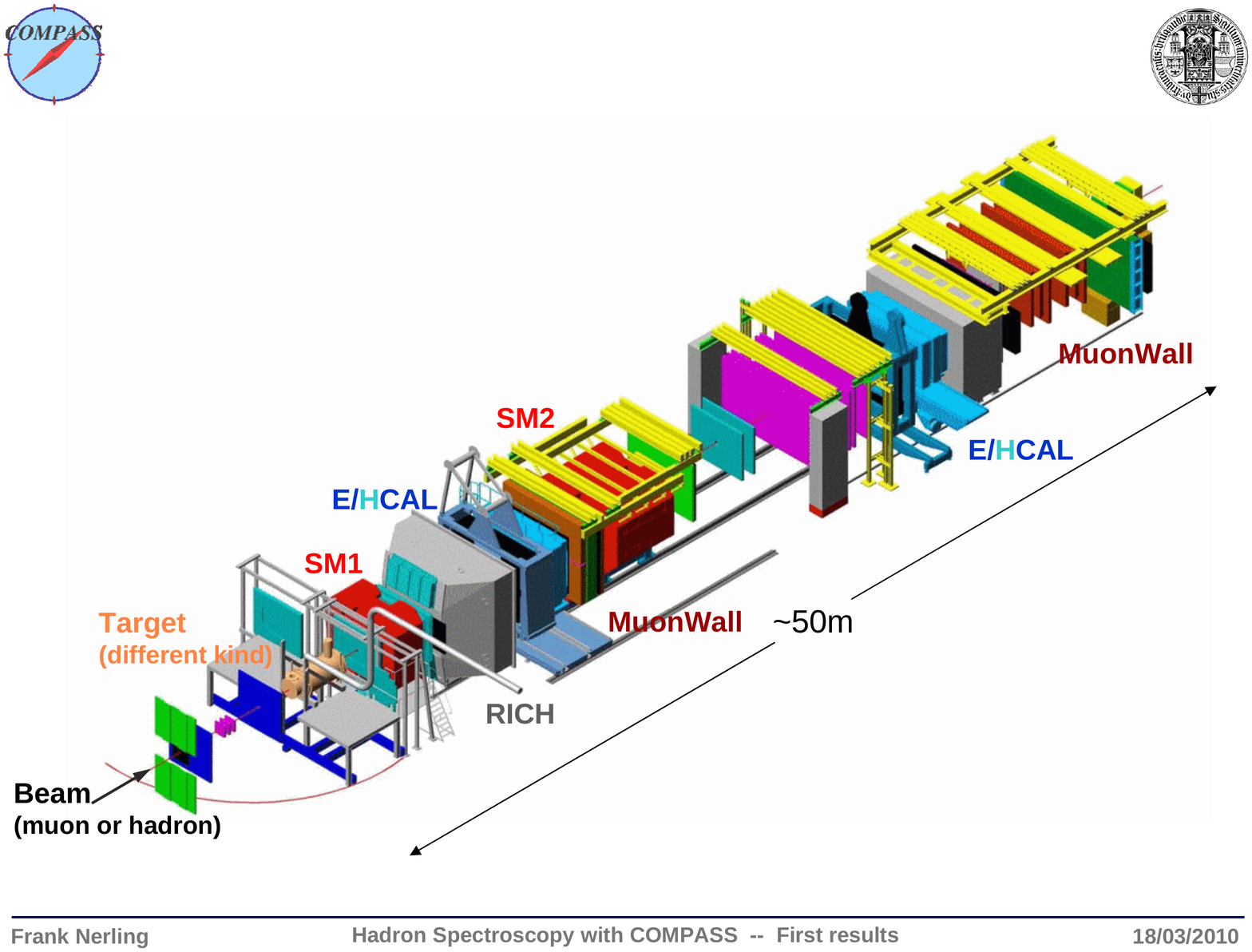}
    \end{center}
  \end{minipage}
  \hfill
  \begin{minipage}[h]{.39\textwidth}
    \begin{center}
      \includegraphics[clip,trim= 20 0 0 0,width=0.8\linewidth,
       angle=0]{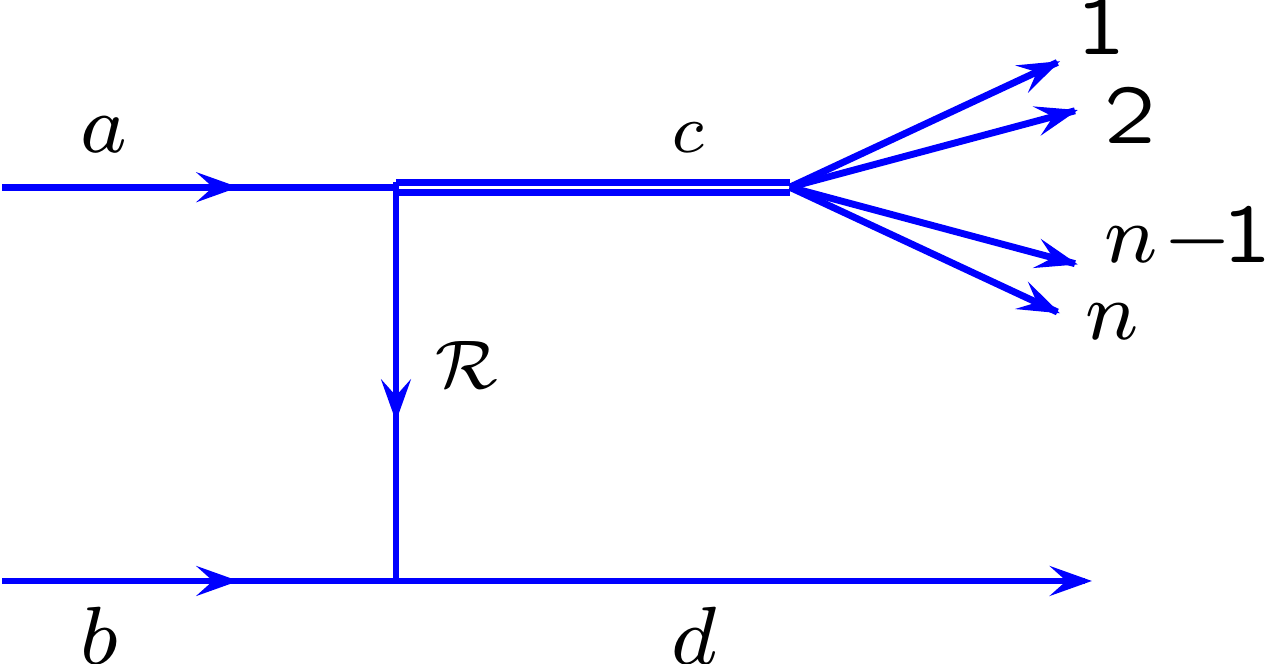}
\vspace{0.3cm}
      \includegraphics[clip,trim= 17 10 3 -10,width=0.8\linewidth,
     angle=0]{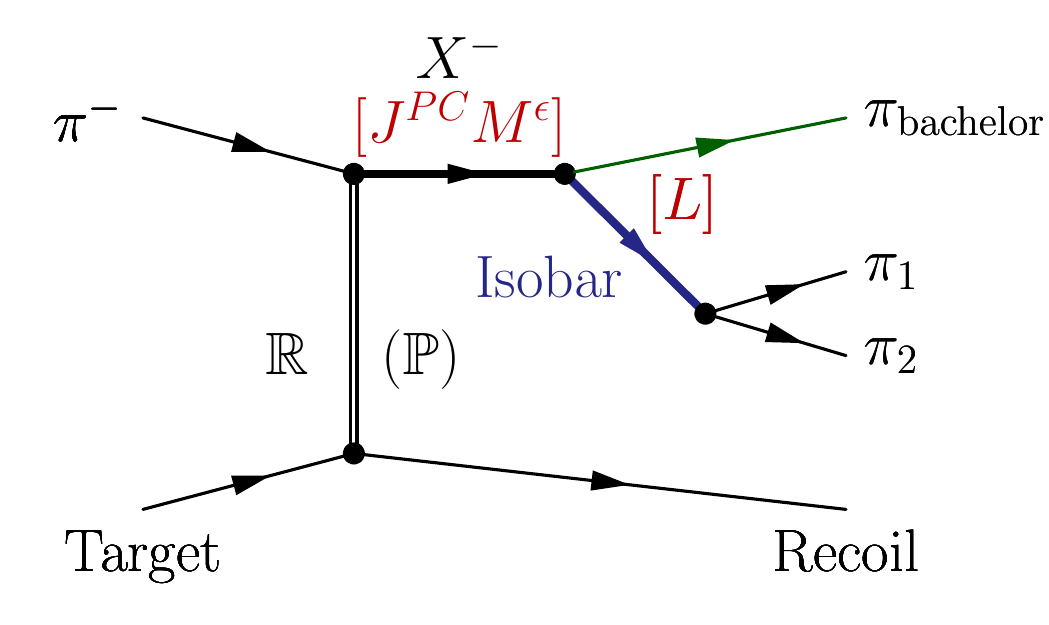}
    \end{center}
  \end{minipage}
      \caption{Left: Sketch of the two-stage COMPASS spectrometer.
               Right: \textit{(top)} Diffractive meson production via $t$-channel Reggeon exchange: The beam particle $a$ is excited to 
                     a resonance $c$ subsequently decaying into $n$ mesons, the target stays intact. 
                     \textit{(bottom)} Diffractive dissociation into 3$\pi$ final states as described in the isobar model: The produced
                     resonance $X^{-}$ with quantum numbers $J^{PC}M^\epsilon$ decays into an isobar with spin $S$ and relative orbital
                     angular momentum $L$ with respect to the $\pi_{\rm bachelor}$. The isobar subsequently decays into two pions.
                     At high energies, the Pomeron is the dominant Regge-trajectory.~~~~~~~~~~~~~~~~~~~~~~~~~~~~~~~~~~~~~~~~~~~~~~~~~~~~~~~~~~~~~~~~~~~~~~~~~~}
       \label{fig:diffrProd_Spectro} 
\end{figure}

\vspace{-0.3cm}
A detailed description of the COMPASS two-stage spectrometer, Fig.\,\ref{fig:diffrProd_Spectro} (left), dedicated to a variety 
of fixed-target physics programmes can be found in \cite{compass:2007}. For the measurements with hadron beams started in 2008, 
a 40\,cm long liquid hydrogen target,
and simple disks of solid material (part of 2009 run) have been 
used. The spectrometer features electromagnetic and hadronic calorimetry in both stages (E/HCAL in Fig.\,\ref{fig:diffrProd_Spectro}). 
Photon detection in a wide angular range with high resolution is crucial for decay channels involving $\pi^{0}$, $\eta$ or $\eta'$.
Nearly 4$\pi$ coverage is achieved for charged and neutral particles in the final states with forward kinematics. 
A Recoil Proton Detector (RPD) consisting of two concentric barrels of scintillator slats 
was introduced to trigger on interactions inside the target 
It detects the recoil particle to ensure exclusivity by a time-of-flight measurement at high accuracy of $\sim$350\,ps. 
%
%
\vspace{-0.3cm}
\section{Analysis of 3$\pi$ final states from diffractive dissociation}
\label{sec.analysis}
\vspace{-0.3cm}
Consider the reaction $a+b \rightarrow c+d$, with $c\rightarrow 1+2+...+n$, where $a$ is the incoming 
beam particle, $b$ the target, $c$ the diffractively produced object decaying into $n$ particles, and $d$ the target 
recoil particle staying intact, see Fig.\,\ref{fig:diffrProd_Spectro} {\it (right/top)}. 
The reaction is described by two kinematic variables, the square of the total 
centre of mass energy $s=(p_a+p_b)^2$ and $t'=|t|-|t|_{\rm min} \ge 0$, with $t=(p_a-p_c)^2$ the squared momentum transfer 
from the target to the beam particle, and $|t|_{\rm min}$ is the minimum absolute value of $t$ allowed by kinematics for a given 
mass $m_c$. 
The basic selection of diffractive 3$\pi$ events common to all analyses discussed in this paper requires exactly one primary 
vertex within the target volumes. One incoming beam particle plus three outgoing charged tracks (charged mode) and one outgoing 
negatively charged track plus exactly 4 clusters detected in the ECALs (neutral mode) are requested, respectively, both with total charge $-1$. 
An exclusivity cut, taking into account the momentum transfer $t'$ to the target, 
ensures that the total energy of the three outgoing pions add up to the beam energy. The partial wave analysis (PWA) is performed
in the range of $0.1 < t' < 1.0$\,GeV$^2$/c$^2$ in order to stay above hardware thresholds and to ensure 
diffractive processes. In case of 2008 data, a cut on the collinearity of the recoil proton 
and the outgoing pion system further suppresses non-exclusive background. 
%
%
\vspace{-0.3cm}
\subsection{Observation of a $J^{PC} = 1^{-+}$ exotic resonance -- 2004 pilot run data}
\label{subsec.2004}
\vspace{-0.25cm}
The PWA is based on the isobar model, see Fig.\ref{fig:diffrProd_Spectro} {\it (right/bottom)}.
A partial waves 
is defined by a set of quantum numbers $J^{PC}M^\epsilon[isobar]L$, with spin $J$, parity $P$ and $C$-parity 
of the resonance $X^{-}$. $M$ and $\epsilon$ (reflectivity) define the spin projection. The resonance decays via an intermediate 
di-pion resonance (the isobar) accompanied by a so-called bachelor pion, with relative orbital angular momentum $L$. We perform our PWA in two steps, a mass-independent fit 
and a subsequent mass-dependent fit. The former is performed on the data binned into 40\,MeV/c$^2$ wide mass intervals, so that no 
assumption on the resonance structure of the $3\pi$ system is made. A total set of 42 waves including a flat background wave is 
fitted to the data using an extended maximum likelihood method, which comprises acceptance corrections. 
The mass-dependent fit is applied to the six main waves, which result from the first step, and uses a $\chi^2$ minimisation. 
The mass dependence is parameterised by relativistic Breit-Wigners (BW) and coherent background if present. 
The employed parameterisation of the spin density matrix has a rank of two, accounting for spin-flip and spin-non-flip amplitudes 
at the baryon vertex. 
\begin{figure}[tp!]
\vspace{-0.6cm}
  \begin{minipage}[h]{.49\textwidth}
    \begin{center}
      \includegraphics[clip,trim= 3 4 22 5,width=0.9\linewidth,
       angle=0]{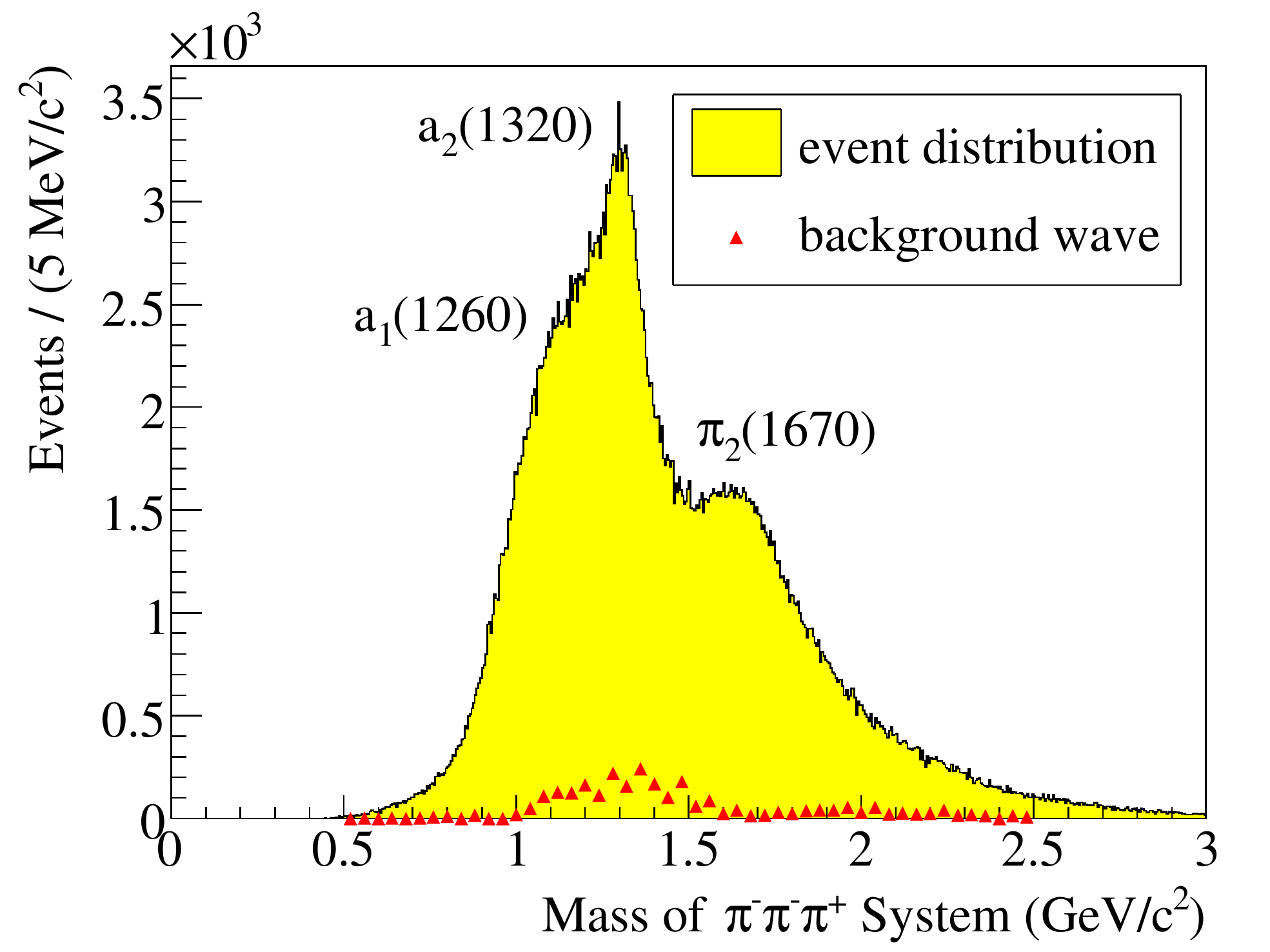}
    \end{center}
  \end{minipage}
  \hfill
  \begin{minipage}[h]{.49\textwidth}
    \begin{center}
      \includegraphics[clip,trim= 24 15 10 360,width=0.95\linewidth,
     angle=0]{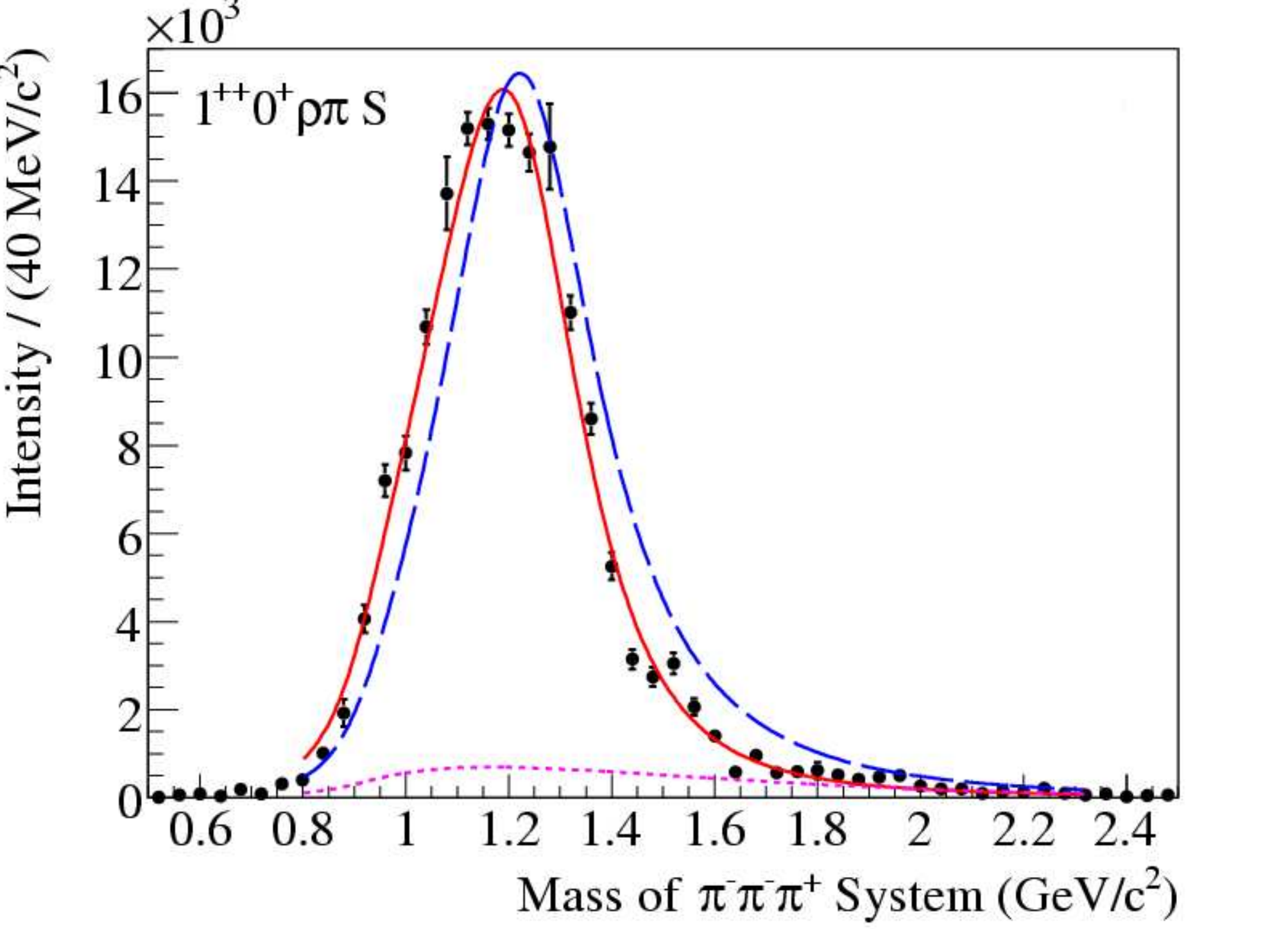}
    \end{center}
  \end{minipage}
\begin{minipage}[h]{.49\textwidth}
    \begin{center}
      \includegraphics[clip,trim= 4 15 30 360,width=0.95\linewidth,
	angle=0]{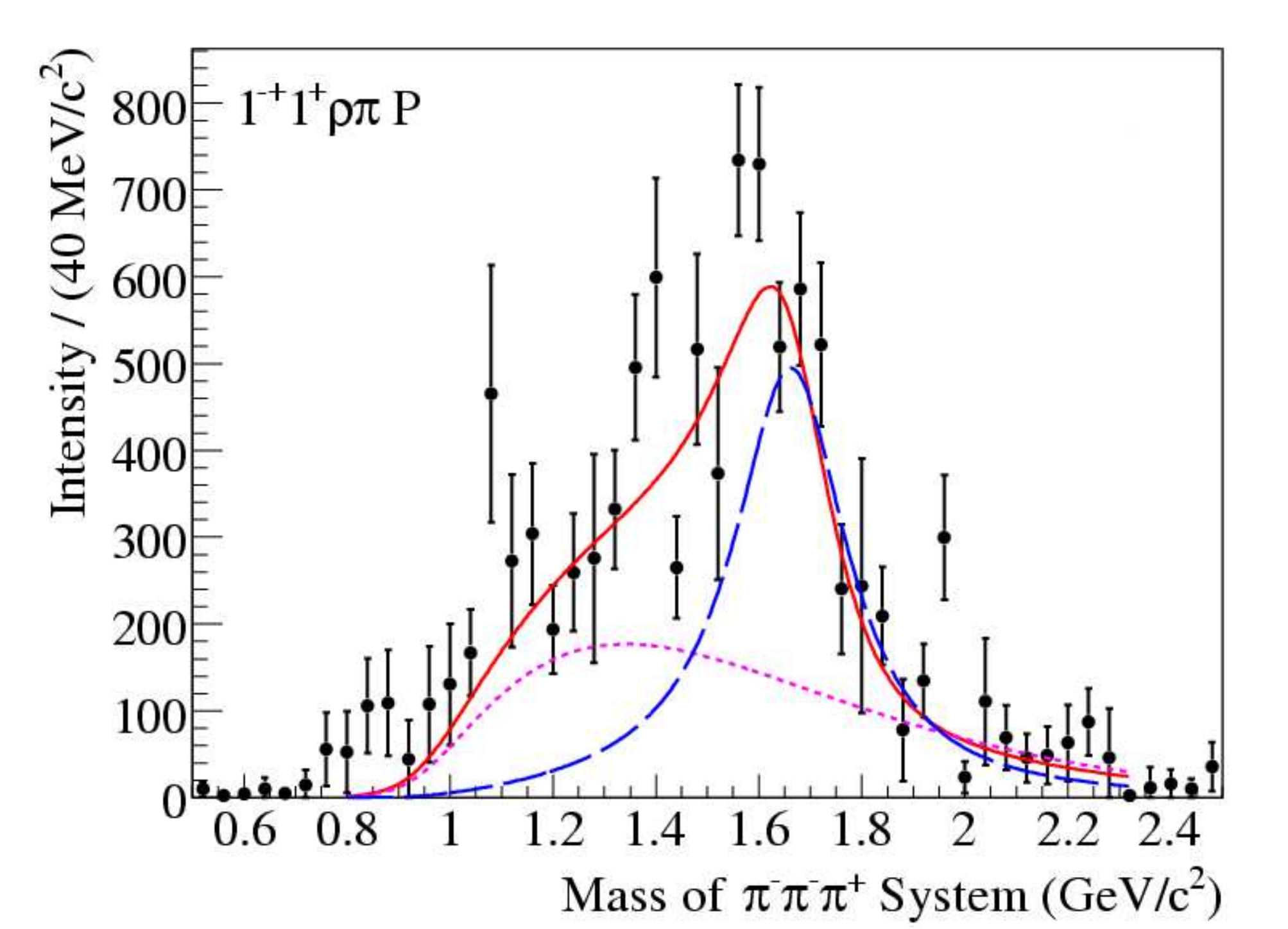}
    \end{center}
  \end{minipage}
  \hfill
  \begin{minipage}[h]{.49\textwidth}
    \begin{center}
      \includegraphics[clip,trim= 24 15 10 360,width=0.95\linewidth,
     angle=0]{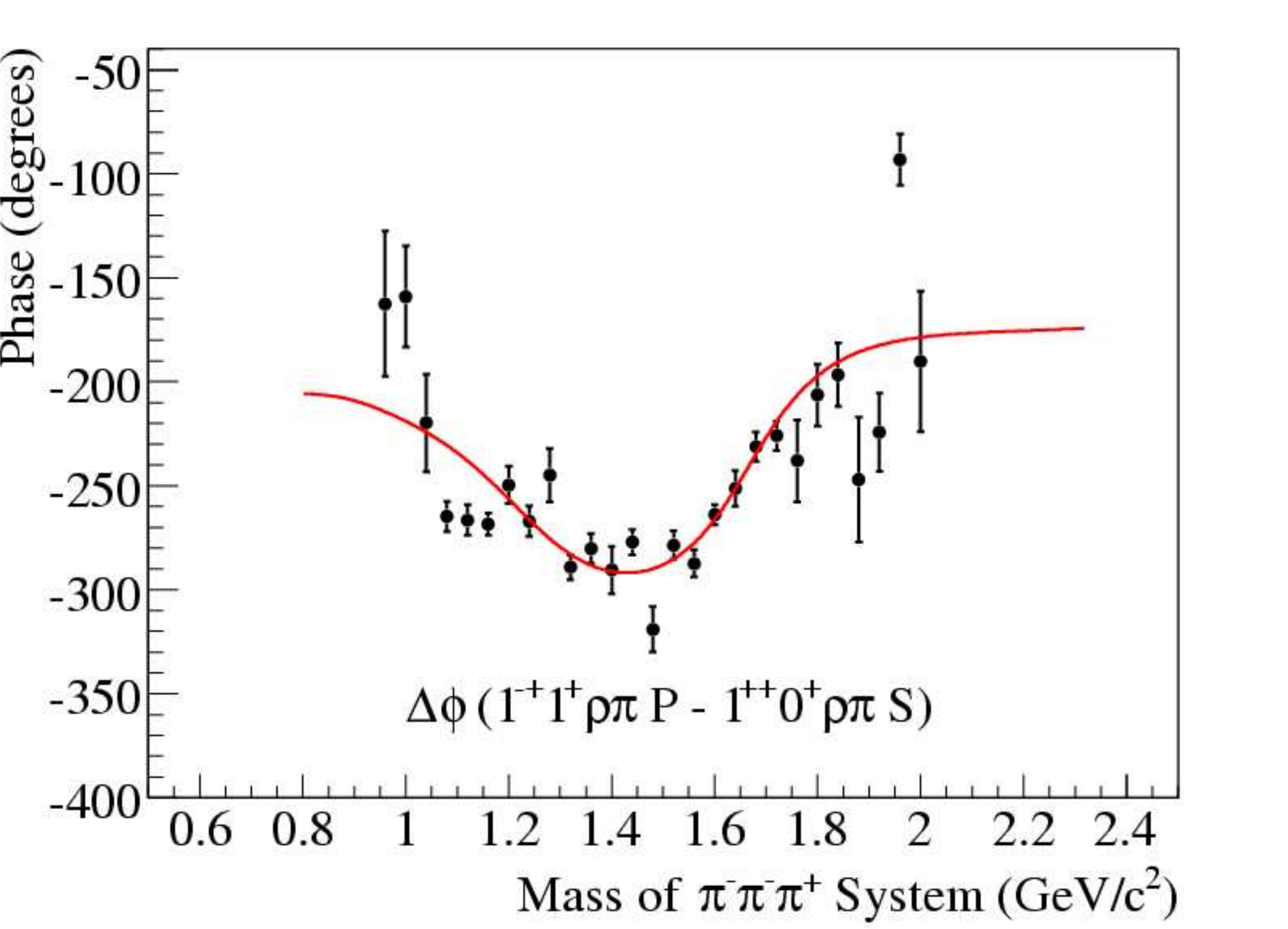}
    \end{center}
  \end{minipage}
      \caption{Top: {\it (Left)} Invariant mass of $\pi^{-}\pi^{+}\pi^{-}$ showing the most prominent resonances (2004 data). {\it (Right)} Example PWA fit for $a_{\rm 1}(1260)$: Intensity of $1^{\rm ++} 0^{\rm +}[\rho^{-} \pi] S$ wave. Bottom: {\it (Left)} Fitted intensity of the exotic $1^{\rm -+} 1^{\rm +}[\rho \pi] P$ wave. {\it (Right)} Phase differences of the exotic $1^{\rm -+}1^{\rm +}$ and the $1^{\rm ++} 0^{\rm +}$ wave -- for details see text.~~~~~~~}
      \label{fig:PWA2004}
\end{figure}
Fig.\,\ref{fig:PWA2004} shows the intensity of the $1^{++}$ wave with the well-established $a_1(1260)$ and that of the spin-exotic $1^{-+}$ wave 
as well as the phase difference $\Delta\Phi$ between the two, for the fits of all six waves, see \cite{Alekseev:2009a}. The black data points represent 
the mass-independent fit, whereas the mass-independent one is overlayed as red solid line, the separation into background 
(purple) and BW (blue) is plotted where applicable. In previous observations, especially the resonant nature of the exotic $1^{-+}1^{+}[\rho\pi] P$ wave is questioned, whereas our data shows a clear and rapid phase motion.
Our result of a mass of $1660\pm 10^{+0}_{-64}$\,MeV/c$^2$ and a width of $269\pm 21^{+42}_{-64}$\,MeV/c$^2$ is consistent with the 
famous $\pi_1(1600)$~\cite{Alekseev:2009a} already reported by previous experiments but still controversially discussed.
%
%
\vspace{-0.3cm}
\subsection{First comparison of neutral versus charged mode  -- 2008 data}
\label{subsec.2008}
\begin{figure}[tp!]
\vspace{-0.6cm}
  \begin{minipage}[h]{.53\textwidth}
    \begin{center}
      \includegraphics[clip,trim= 3 26 10 22,width=0.65\linewidth,
       angle=90]{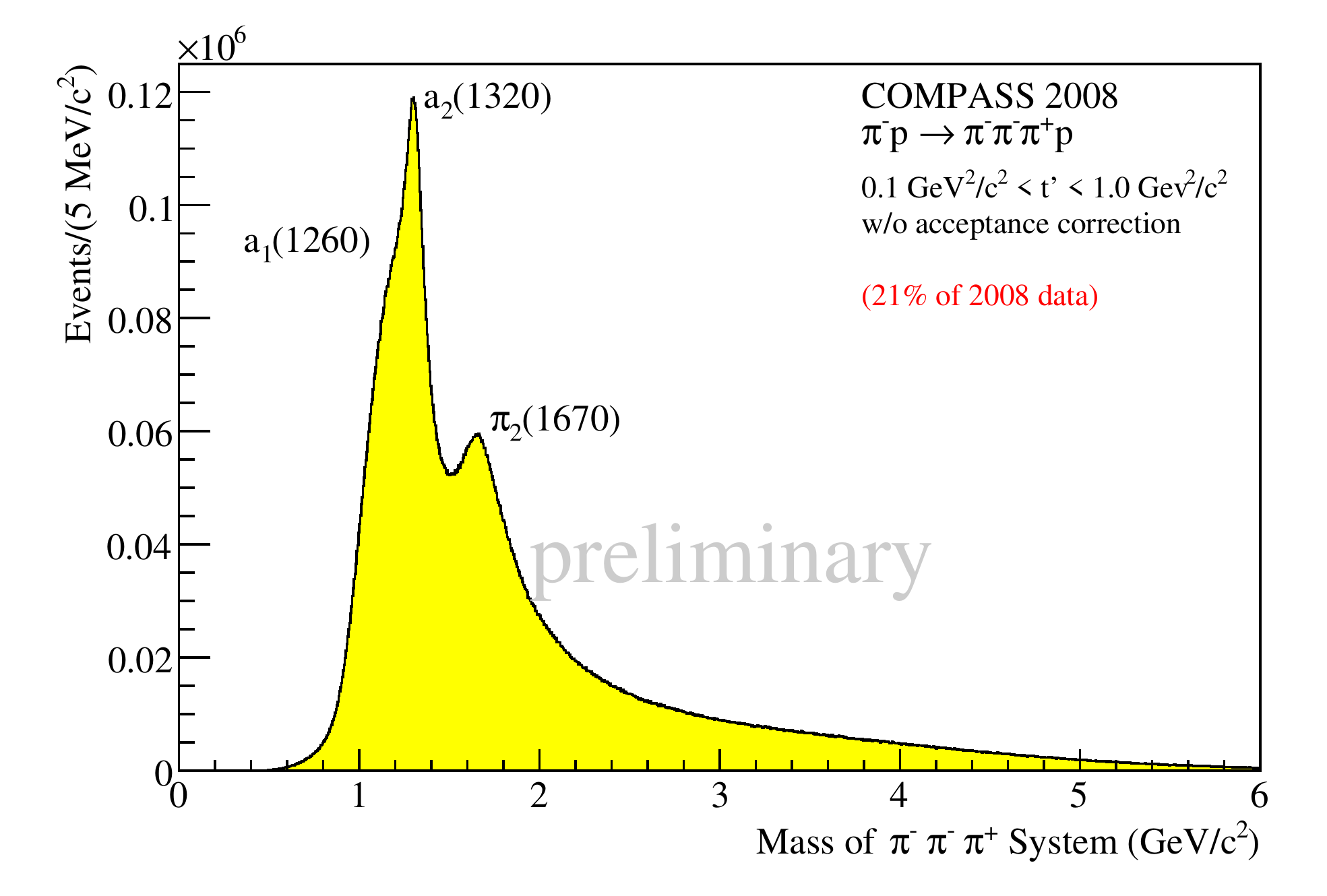}
    \end{center}
  \end{minipage}
  \hfill
  \begin{minipage}[h]{.45\textwidth}
    \begin{center}
      \includegraphics[clip,trim= 0 8 26 4,width=0.85\linewidth,
     angle=90]{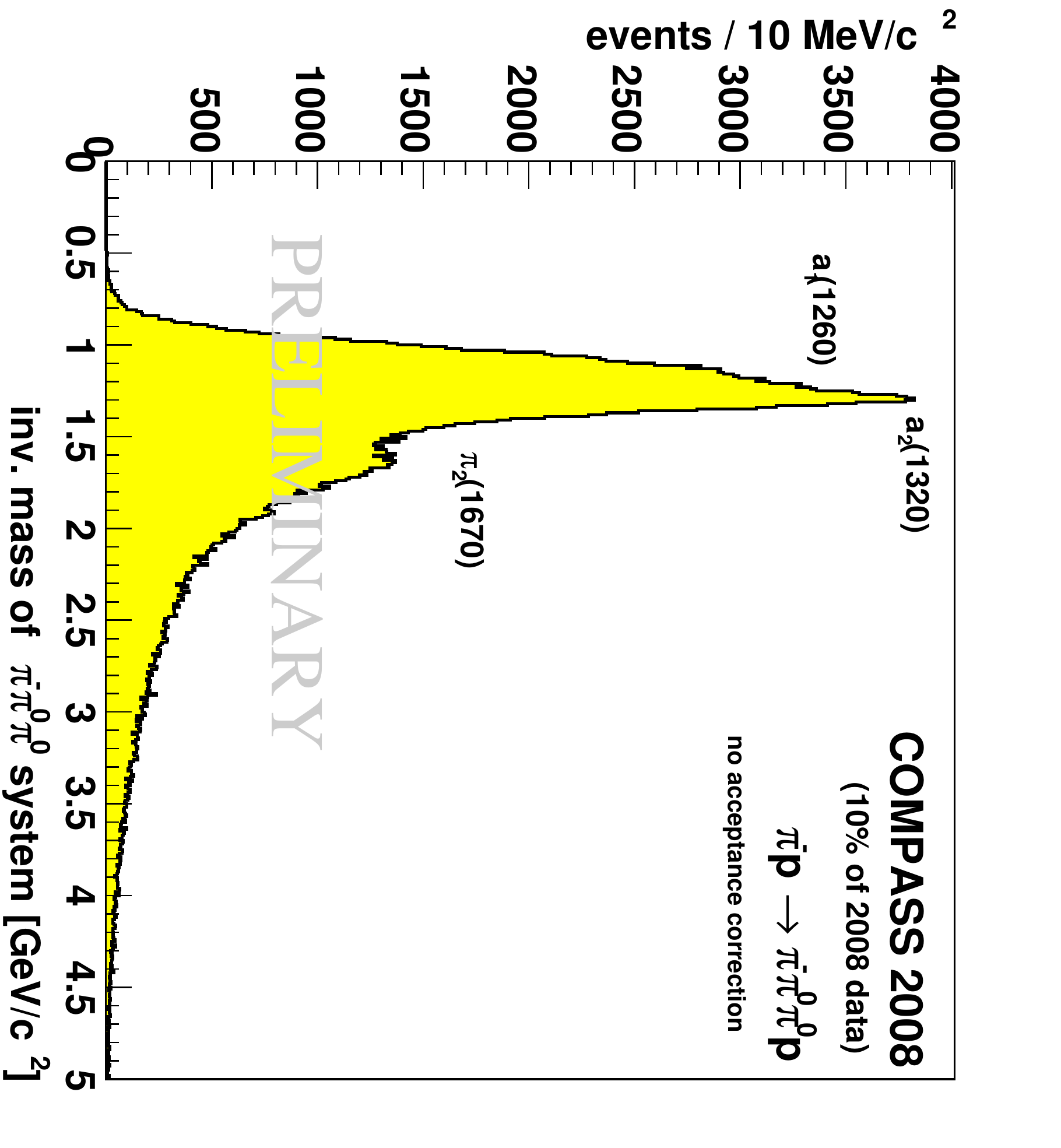}
    \end{center}
\vspace{-0.4cm}
  \end{minipage}
\vspace{-0.2cm}
  \begin{minipage}[h]{.49\textwidth}
    \begin{center}
      \includegraphics[clip,trim= 3 0 22 9,width=0.9\linewidth,
	angle=0]{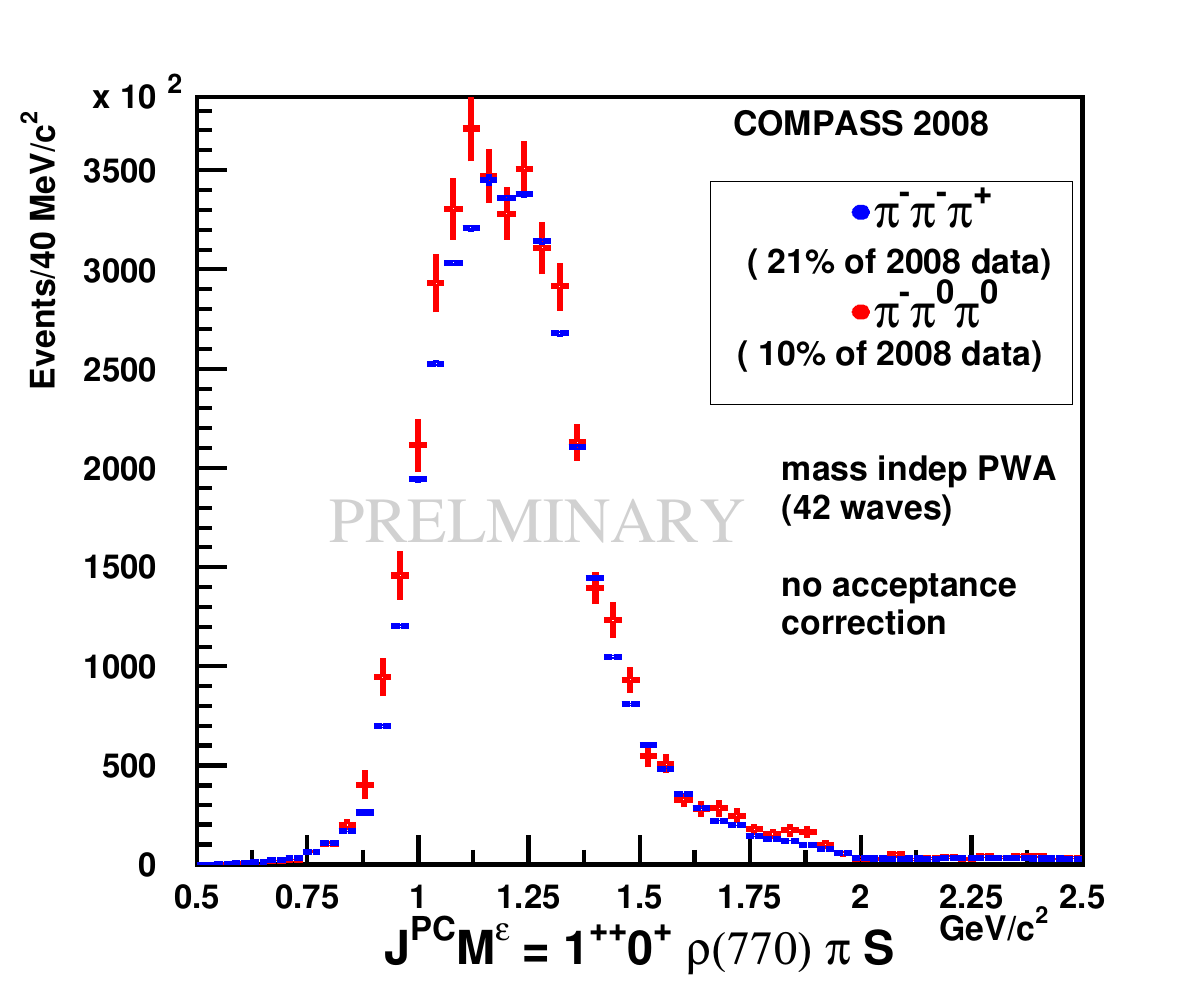}
    \end{center}
  \end{minipage}
  \hfill
  \begin{minipage}[h]{.49\textwidth}
    \begin{center}
      \includegraphics[clip,trim= 3 0 22 9,width=0.9\linewidth,
     angle=0]{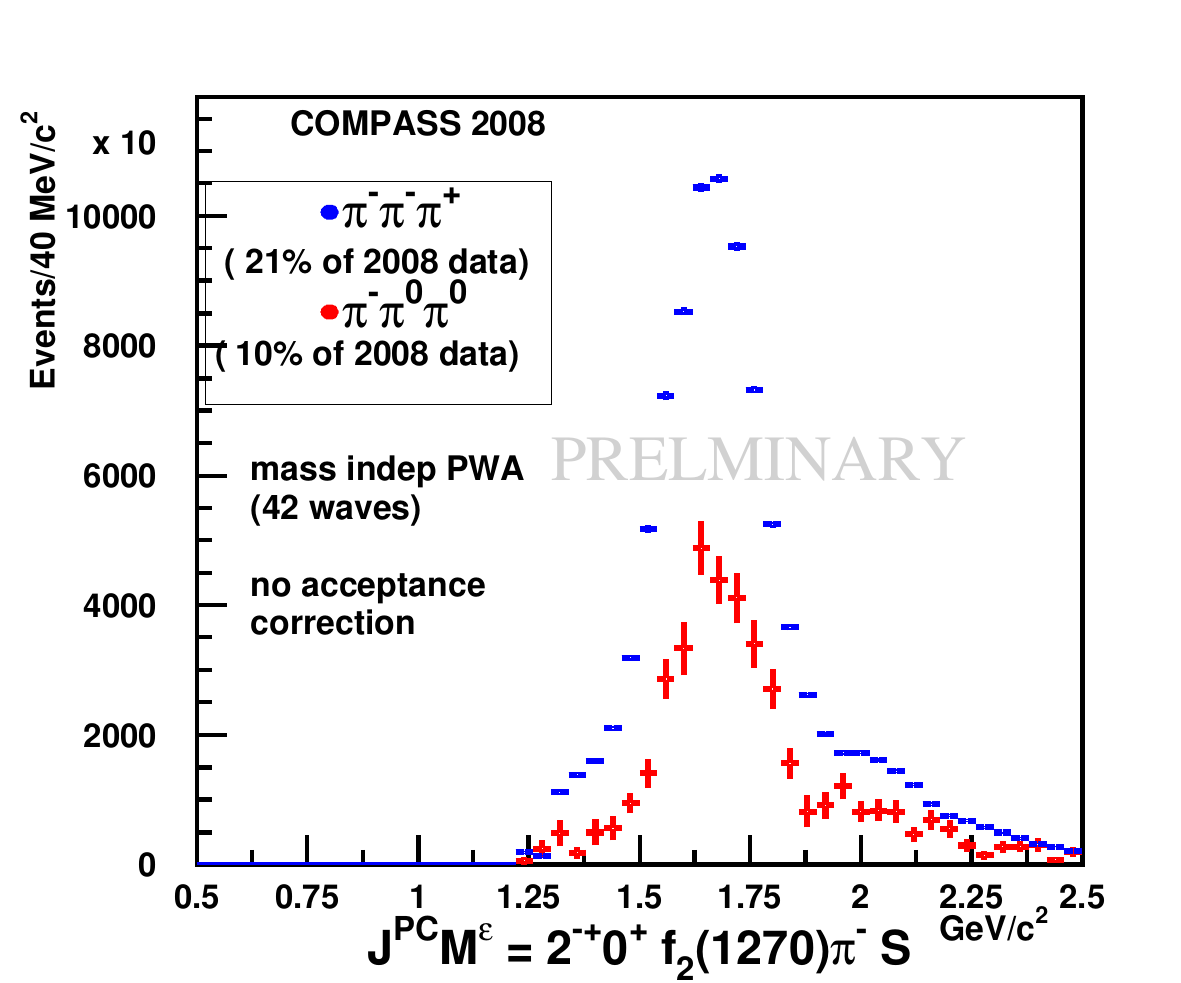}
    \end{center}
  \end{minipage}
      \caption{Comparison of 3$\pi$ analyses neutral vs. charged mode (2008 data). 
	Top: Invariant mass of $\pi^{-}\pi^{+}\pi^{-}$ 
	and $\pi^{-}\pi^{0}\pi^{0}$ 
	showing the most prominent resonances. Bottom: Exemplary intensities of main waves, for comparison, neutral and charge mode intensities were normalised to the well-established $a_{\rm 2}(1320)$: $2^{\rm ++} 1^{\rm +}[\rho\pi]D$ wave. {\it (Left)} $a_{\rm 1}(1260)$: 
	$1^{\rm ++} 0^{\rm +}[\rho\pi]S$ wave, 
	{\it (Right)} $\pi_{\rm 2}(1670)$: $2^{\rm -+} 0^{\rm +}[f_{\rm 2}(1270)\pi]S$ wave.~~~~~~~~~~~~~~~~~~~~~~~~~~~~~~~~~~~~~~~~~~~~~~~~~~~~}
      \label{fig:PWA2008}
\end{figure}

\vspace{-0.25cm}
The event selections have been discussed more detailed \cite{nerling:2009,haas:2009}.
The resultant invariant mass spectra for charged and neutral mode are shown in Fig.\,\ref{fig:PWA2008} {\it (top)}, exhibiting the 
same prominent resonances as in Fig.\,\ref{fig:PWA2004}. In this first analysis, we apply the same PWA model as for the 2004 data. 
The mass-independent fit results are compared in Fig.\,\ref{fig:PWA2008} {\it (bottom)}, normalised using the $a_2(1320)$ to 
account for different statistics. 
Even though acceptance corrections have not yet been applied (estimated effect up to 20\%), the comparison is already quite promising. 
From simple isospin symmetry consideration, we expect for all isobars decaying into $\rho\pi$ 
the same intensities for both modes, whereas for those going to $f_2\pi$, a suppression factor of two is expected for the neutral mode, 
simply due to Clebsch-Gordon coefficients. Indeed Bose-Symmetrisation with the bachelor pion might modify the picture. We checked, however, 
that the effect is the same for $\rho\pi$ decays and thus cancels out, whereas for $f_{0,2}\pi$ channels, the effects can be large in 
general, but no effect was found for the one chosen here. 
Indeed similar intensities are found for the $1^{++}0^{+}[\rho(770)\pi]S$ wave in both modes, and about half intensity in the 
$2^{-+}0^{+}[f_2(1270)\pi]S$ wave for the neutral mode -- as expected. 
Such kind of isospin-symmetry checks will validate independent confirmation in case of new states, like e.g. the $\pi_1(1600)$, simultaneously observed in different decay modes.
\vspace{-0.4cm}
\section*{Summary}
\vspace{-0.3cm}
First results on the COMPASS hadron data show the high potential for meson spectroscopy. Our 2004 pilot run 
data show a significant spin exotic signal consistent with the controversially discussed $\pi_1(1600)$. 
The data collected in 2008/09 exceeds the world statistics by up to two orders of magnitude. One advantage of COMPASS 
as compared to previous fixed-target experiments is the ability for detecting final states involving charged and neutral particles. 
COMPASS has not only access to many decay channels but also to higher masses exceeding 3.5\,GeV/c$^2$.
\vspace{-0.4cm}
\section*{Acknowledgements}
\vspace{-0.3cm}
This work is supported by the BMBF (Germany), particularly the ``Nutzungsinitiative CERN''.
\vspace{-0.8cm}
\section*{References}
\vspace{-0.4cm}

\begin{thebibliography}{99}
\bibitem{compass:2007} P.~Abbon {\it et al.}, COMPASS collaboration, \Journal{Nucl. Instrum. Meth.}{A577}{455}{2007}.
\bibitem{Morningstar:2004} K.J.~Juge, J.~Kuti, C.~Morningstar, \Journal{AIP Conf. Proc.}{688}{193}{2004}.
\bibitem{Adams:1998} G.S.~Adams {\it et al.}, \Journal{Phys. Rev. Lett.}{81}{5760}{1998}.
\bibitem{Khokhlov:2000} Y.~Khokhlov, \Journal{Nucl. Phys. A}{663}{596}{2000}.
\bibitem{Amelin:2005} D.V.~Amelin {\it et al.}, \Journal{Phys. Atom. Nucl.}{68}{359}{2005}.
\bibitem{Dzierba:2006} A.R.~Dzierba {\it et al.}, \Journal{Phys. Rev. D}{73}{072001}{2006}.
\bibitem{Alekseev:2009a} M.~Alekseev {\it et al.}, COMPASS collab., subm.\,to Phys.\,Rev.\,Lett, arXiv:0910.5842v2 (2010).
\bibitem{nerling:2009} F.~Nerling {\it et al.}, COMPASS collab., AIP Conf. Proc. (2009), to\,be\,published.
\bibitem{haas:2009} F.~Haas {\it et al.}, COMPASS collab., AIP Conf. Proc. (2009), to\,be\,published.
\end{thebibliography}

\end{document}